\documentclass[conference]{IEEEtran}
\IEEEoverridecommandlockouts

\usepackage{cite}

\usepackage{siunitx}
\usepackage{graphicx}
\usepackage{subcaption}
\usepackage{mathtools}
\usepackage{orcidlink}

\def\BibTeX{{\rm B\kern-.05em{\sc i\kern-.025em b}\kern-.08em
    T\kern-.1667em\lower.7ex\hbox{E}\kern-.125emX}}
\begin{document}

\title{Modeling the Potential of Message-Free Communication via CXL.mem
\thanks{
This work has been accepted for publication in SCA/HPCAsia 2026: Supercomputing Asia and International Conference on High Performance Computing in Asia Pacific Region (SCA/HPCAsia 2026).

This work was supported by the Plasma-PEPSC project, which has received funding from the European High-Performance Computing Joint Undertaking (JU) as well as the German Federal Ministry of Research, Technology, and Space (BMFTR) through Grant Agreement No. 101093261.
Part of the performance results have been obtained on systems in the test environment BEAST (Bavarian Energy Architecture \& Software Testbed) at the Leibniz Supercomputing Centre.
Grok 3 and Copilot (GPT 4.1) were used to help improve text quality, spelling, and grammar.}
}

\makeatletter
\newcommand{\linebreakand}{%
  \end{@IEEEauthorhalign}
  \hfill\mbox{}\par
  \mbox{}\hfill\begin{@IEEEauthorhalign}
}
\author{
\IEEEauthorblockN{1\textsuperscript{st} Stepan Vanecek}
\IEEEauthorblockA{\textit{Technical University of Munich} \\
    Garching, Germany \\
    stepan.vanecek@tum.de\,\orcidlink{0009-0008-4120-9472}} 
\and
\IEEEauthorblockN{2\textsuperscript{nd} Matthew Turner}
\IEEEauthorblockA{\textit{Hewlett Packard Enterprise} \\
    Houston, TX, USA \\
    matthew.turner@hpe.com\,\orcidlink{0000-0001-8529-6709}} 
\and
\IEEEauthorblockN{3\textsuperscript{rd} Manisha Gajbe}
\IEEEauthorblockA{\textit{} 
    Folsom, CA, USA \\
    manisha.gajbe@gmail.com\,\orcidlink{0009-0004-6589-4209}} 
\linebreakand
\IEEEauthorblockN{4\textsuperscript{th} Matthew Wolf}
\IEEEauthorblockA{\textit{} 
    Portland, OR, USA\\
    wolf.m7@gmail.com\,\orcidlink{0000-0002-8393-4436}} 
\and
\IEEEauthorblockN{5\textsuperscript{th} Martin Schulz}
\IEEEauthorblockA{\textit{Technical University of Munich} \\
    Garching, Germany \\
    schulzm@in.tum.de\,\orcidlink{0000-0001-9013-435X}} 
}

\maketitle

\begin{abstract}
Heterogeneous memory technologies are increasingly important instruments in addressing the memory wall in HPC systems. While most are deployed in single node setups, CXL.mem is a technology that implements memories that can be attached to multiple nodes simultaneously, enabling shared memory pooling. This opens new possibilities, particularly for efficient inter-node communication.

In this paper, we present a novel performance evaluation toolchain combined with an extended performance model for message-based communication, which can be used to predict potential performance benefits from using CXL.mem for data exchange.
Our approach analyzes data access patterns of MPI applications: it analyzes on-node accesses to/from MPI buffers, as well as cross-node MPI traffic to gather a full understanding of the impact of memory performance.
We combine this data in an extended performance model to predict which data transfers could benefit from direct CXL.mem implementations as compared to traditional MPI messages. 
Our model works on a per-MPI call granularity, allowing the identification and later optimizations of those MPI invocations in the code with the highest potential for speedup by using CXL.mem.

For our toolchain, we extend the memory trace sampling tool Mitos and use it to extract data access behavior.
In the post-processing step, the raw data is automatically analyzed to provide performance models for each individual MPI call.
We validate the models on two sample applications -- a 2D heat transfer miniapp and the HPCG benchmark -- and use them to demonstrate their support for targeted optimizations by integrating CXL.mem.
\end{abstract}


%%%%%%%%%%%%%%%%%%%%%%%%%%%%%%%%%%%%%%%%%%%%%%%%%%%%%%%%%%%%%%%%%%%%%%%%%%%%%%
\section{Motivation and Contributions}\label{sec:motivation}

Modern high-performance systems in HPC, AI, and cloud are facing the challenge of hitting the memory wall with traditional DRAM-based memory solutions~\cite{wulf1995hitting}.
Various new technologies addressing this challenge have emerged in recent years to streamline data accesses.
New manufacturing approaches, such as die stacking with High-Bandwidth Memory (HBM), show significant advances~\cite{black2013stacking}.
However, due to their higher price and limited capacity, we often only consider them in a heterogeneous memory setup, used in conjunction with traditional DDR DRAM~\cite{meswani2015heterogeneous}. %DDR DRAM~\cite{meswani2015heterogeneous,hagleitner2023architecture}
Intel's now-discontinued non-volatile Optane memory was used as a larger but slower complement to DDR DRAM (and/or HBM)~\cite{hady2017platform}.
One of the latest developments is memory attached to one or more nodes via the Compute Express Link (CXL), essentially forming a heterogeneous memory setup with the base (DRAM) memory.

CXL.mem (2.0+) enables not only attaching memory directly on-node (scale-up), but also off-node (scale-out), implicitly enabling high-bandwidth, low-latency data transfers between the host and the detachable memory device~\cite{sun2023demystifying,sharma2022compute}.
With CXL.mem 3.0, coming systems will further be able to pool these devices, creating a cache-coherent memory segment that allows multiple hosts (or nodes) to read and write from/to the same memory pool~\cite{CXL32}. %~\cite{cxl3x}.
This positions CXL.mem as a broker in cross-node data communication -- similar to shared memory buffers for on-node communication -- and potentially establishes a new architectural paradigm between classic shared memory and shared nothing systems. 
The question now is how to best use this new capability, which we refer to as message-free communication, and whether it can or should replace current communication mechanisms via message passing, which in HPC typically means MPI-based programs with direct copies.
Several studies (\cite{tran2024omb,ahn2024mpi}) simulate MPI communication over shared CXL buffers, reporting up to 42x speedup, which demonstrates the potential of this approach.

These studies evaluate only the data transfer, and do not consider the application context or the implications of the MPI buffer allocation.
To obtain actionable insights for real applications, we require a detailed understanding of MPI buffer-related data movements, combined with an extended performance model that predicts the performance of message-free communication via direct CXL.mem buffer access.
The model must account for (1) CPU–memory load/store operations and (2) inter-node data transfers, whether via a network or directly through CXL.mem.
This combined view will allow us to identify which communications will benefit from direct CXL.mem communication and which should remain as standard network transfers.

Many popular performance monitoring and analysis tools aim at analyzing both on-node and cross-node performance; examples include HPCToolkit, Tau, Extrae, and Paraver.
However, none of these tools focus their analysis specifically on the data access patterns of MPI buffers as a critical piece of an application's memory space especially in the context of evaluating CXL.mem as an alternative MPI buffer allocation.

In this paper, we fill this gap and present a novel toolchain that 1) collects and analyzes traces and metrics covering the buffer accesses in existing MPI implementations and 2) uses this information to predict the performance of a future CXL.mem-based implementation using a message-free approach, separately for each MPI call\footnote{We only consider MPI send-recv communication in this study, but the principles apply to other communication types as well, including collectives.}.
Our toolchain builds on and extends the Mitos tool: in particular, we extend Mitos to work seamlessly with parallel MPI codes, allowing for analysis of parallel applications. 
It further collects MPI traces to connect them with the data access context to minimize user effort to generate the predictions.

The per-MPI-call granularity enables direct evaluation of the potential benefit of refactoring particular MPI operations in the source code towards a direct CXL-based data exchange.
This helps future users identify key optimization opportunities related to optimizing their codes on those new system architectures:
\begin{enumerate}
    \item In which cases will MPI baseline perform better, and which MPI calls will benefit from CXL?
    \item By replacing which MPI calls can I expect to get the most performance gain? (Where do I invest my time first?)
    \item If faced with limited capacity on CXL, which buffers should be prioritized?
\end{enumerate}
These questions are essential to aim developers' focus towards particular data exchanges, maximizing the performance and minimizing the effort for optimizing user applications for these specific system architectures.
From another perspective, our toolchain can be used for better system design: which performance parameters do my memory pooling devices need in order to offer better performance than MPI communication for a target set of applications?

After presenting our data collection tool and the performance model, we demonstrate its validity in two use cases -- a 2D heat transfer code and the HPCG benchmark.
Given the unavailability of CXL.mem 3.0+ products on the market\footnote{First ones, such as Intel Diamond Rapids, are expected in 2026.}, we mimic the CXL memory with an on-node setup using heterogeneous DDR+Optane memory.
We validate the predictions against our shared memory buffer reference implementation.
The model clearly identifies the potential gain/loss for different MPI calls, providing user guidance.
Finally, we present the prediction for the target multi-node setup with cross-node communication, estimating up to 1.59x overall speedup compared to the MPI baseline.

%%%%%%%%%%%%%%%%%%%%%%%%%%%%%%%%%%%%%%%%%%%%%%%%%%%%%%%%%%%%%%%%%%%%%%%%%%%%%%
\section{Background}

Our work focuses on leveraging CXL.mem from the CXL protocol family in HPC applications, particularly its use for direct buffer-to-buffer transfers via pooled shared memory -- an approach we call message-free communication.

CXL.mem, built atop PCIe as the physical layer, enables expanding memory through cache-coherent attachment of on- or off-node (scale-up/out) memory segments.
In our study, we consider a setup with off-node Type 3 (CXL 3.0 or later, built atop PCIe 6.0+) disaggregated pooled memory, where one memory segment can be attached to multiple nodes simultaneously and used for direct, message-free data transfers. 
However, its performance characteristics -- both on-node and off-node —- make it unclear when message-free communication should be preferred over traditional messaging.

\subsection{(Expected) Performance of CXL}

CXL.mem has established itself as a very promising technology to optimize cost, performance, and energy demands for different applications in HPC, AI, and the cloud.
Analyses of older CXL performance are already available.
Since CXL 3.0 has yet to enter the market, claims about performance, also used in simulators, have yet to be verified.

Multiple studies explore data allocation policies to maximize application performance~\cite{wang2024exploring,sun2023demystifying}.
Other works~\cite{liu2024dissecting,sun2023demystifying,wahlgren2022evaluating} focus on deepening the understanding of performance impacts of CXL data allocation, focused on CXL 2.0 or earlier.
Studies ~\cite{fridman2023cxl} and~\cite{foyer2023survey} mention ca. \SI{20}{\percent} bandwidth degradation and \SIrange{100}{200}{\nano\second} additional latency with on-node CXL, while CXL 3.2 specification~\cite{CXL32} mentions an expected 10-30\% bandwidth degradation and recommends an \SI{80}{\nano\second} overhead target for reads.
Theoretical CXL memory bandwidth is limited by both the PCIe data transfer speed (\SI{128}{\giga\byte} for PCIe 6.0 x16) and the memory technology itself (from tens of \unit[per-mode = symbol]{\giga\byte\per\second} for low-end DDR4 to hundreds of \unit[per-mode = symbol]{\giga\byte\per\second} for high-end DDR5 theoretical peak BW).
Moving to disaggregated memory systems with switches and complex topologies will further impact both bandwidth and latency however, the extent of this impact will only become clear once the first market products arrive.

\subsection{MPI over CXL}

Several studies have demonstrated the viability and potential of using CXL as a data transfer medium for cross-node MPI communication.
Tran et al.~\cite{tran2024omb} evaluate point-to-point data transfer over CXL, using QEMU-CXL simulator~\cite{qemu_cxl_docs}.
Anh et al.~\cite{ahn2024mpi} extend this model to MPI collectives, modeling (and simulating) up to 42x speedup over traditional \texttt{MPI\_Allgather}, using CXL latency of \SIrange{300}{400}{\nano\second}.
However, these studies focus solely on data transfer, ignoring data access patterns, and evaluate isolated MPI operations rather than full applications.

\subsection{Heterogeneous Memory Modeling and Data Placement}

While heterogeneous memory systems share some properties with NUMA systems in terms of exposing different latencies, and are often managed with OS techniques and abstractions borrowed from NUMA, data placement is vastly different. 
The limitations of first-touch data placement policy in heterogeneous memory systems require new solutions.
The research in this area has mostly revolved around on-node setups (DDR+X).
Nonell et al.~\cite{nonell2018applicability} describe a mechanism for live allocation decisions based on Intel PEBS (Precise event-based sampling).
Others present offline strategies to help the user make optimal allocation decisions~\cite{dulloor2016data,peng2017rthms}.
The H2M library~\cite{klinkenberg2022h2m} facilitates data placement decisions to higher-level users, and Foyer~et~al.~\cite{foyer2023survey} present a survey of available techniques for emulating various heterogeneous memories.

\subsection{Application Performance Analysis}

To understand the performance impact of using CXL.mem as a communication medium, we must create a comprehensive performance prediction toolchain building on data gathered from two sources:
1)~samples of load (LD) accesses to understand the on-node data movement profile, and 2)~MPI traces to gain information about (cross-node) communication.
Many widely used tools for both of these individual tasks are available.
However, since these tasks are tightly coupled, they require a common analysis.

\subsubsection{Mitos and Other Memory-focused Tools}

We build on the Mitos\cite{gimenez2015mitos} tool, originally developed to collect memory sample traces for MemAxes\cite{gimenez2014dissecting}, a performance analysis and visualization tool. We describe Mitos further in Sec.~\ref{sec:mitos}.

Apart from Mitos, vendor-developed tools, like Intel VTune~\cite{intel_vtune} or AMD uProf~\cite{uprof} can capture similar traces, but are closed source and vendor specific.
The Intel PIN~\cite{luk2005pin} is an architecture-independent dynamic instrumentation tool used to intercept operations such as LD/ST. However, its pure software implementation makes it slow, and its raw focus on LD/ST instructions requires substantial work to include realistic cache models.
HPCtoolkit~\cite{10.1145/2555243.2555271} and Tau~\cite{shende2006tau} offer means to collect HW (PEBS) samples, including different views on data accesses.
However, extracting the necessary data and adding our specific features to these feature-rich tools would be significantly more complex, hence we build atop the more lightweight Mitos.
We also mention PAPI~\cite{mucci1999papi} and LIKWID~\cite{gruber2019likwid} -- two widely used tools providing a platform-agnostic abstraction to performance counter collection.
An extension to PEBS sample collection in PAPI has been proposed~\cite{weaver2016advanced} but is even more low-level.
We integrate PAPI into Mitos in our work to provide general application characteristics.

\subsubsection{MPI Communication Tracing/Analysis}

Similarly to the memory-centric analyses, multiple well-established tools help with understanding MPI communication performance.
Prominent examples are the Extrae~\cite{extrae} (instrumentation) and Paraver~\cite{pillet1995paraver} (visualization) or Scorep~\cite{scorep} (profiling and tracing), Scalasca~\cite{geimer2010scalasca}, and Vampir~\cite{knupfer2008vampir} 
(analysis and visualization) toolchains.
They provide a detailed view into the parallel computation and communication performance, essential for optimizing performance, such as for load imbalance or insufficient computation and communication overlap.
While providing invaluable information on communication efficiency, memory analysis is not in their main scope.
Finally, HPCtoolkit, TAU or Intel VTune (mentioned above) also offer MPI tracing functionality, though separated from the memory analysis.

%%%%%%%%%%%%%%%%%%%%%%%%%%%%%%%%%%%%%%%%%%%%%%%%%%%%%%%%%%%%%%%%%%%%%%%%%%%%%%
\section{Mitos Sampling}\label{sec:mitos}

Hardware vendors have introduced various performance-monitoring units (PMUs) in recent years to facilitate understanding of system behavior and performance.
Hardware counters and hardware-based sampling are two main categories.
Sampling is a low-overhead technique, enabled by dedicated circuits, that captures hardware events with high detail.
Various clues regarding the current program and system state are recorded.
It presents a unique opportunity to understand selected events in detail, which enables, in the case of Mitos, providing data address space-, hardware-, and code-related contexts with each observed event.

At a high level, Mitos simplifies hardware event sampling by offering a robust, user-friendly API compared to the complex and poorly documented native interfaces.
Mitos captures \textbf{Intel PEBS} samples, specific to Intel CPUs.\footnote{We extend Mitos to \textbf{AMD IBS} (Instruction Based Sampling) -- both IBS Fetch and IBS Op. However, we do not build on IBS sampling in this study.}
It uses the \texttt{perf\_event\_open} call of the Linux perf API to configure and start the sampling and offers a configurable, vendor-agnostic API for collecting low-level sample-based information.
Mitos provides raw data with contextual metadata, which is key for further analysis.
This distinguishes Mitos from vendor-specific tools, such as Intel PIN tool or Intel VTune.~\cite{gimenez2015mitos}

\subsection{Additional Context Collected in Mitos}

Beyond collecting raw samples, Mitos captures additional context, linking each sample to the underlying hardware topology, source code, and, optionally, problem-domain dimensions contextx.
The first key addition is a hardware topology snapshot, collected via the hwloc API~\cite{hwloc}.
This lays out details on processor cores, caches, and NUMA regions, essential for understanding data flows.

Mitos also records the instruction pointer (IP) for each sample\footnote{Hardware-supported sampling minimizes skid (although it may be non-zero).~\cite{gottschall2021tip}}, enabling attribution to source code and associated data structures.
Together with the base load address\footnote{The difference between the address in the ELF file and the addresses in memory.~\cite{glibc}} 
(recorded by Mitos), Dyninst~\cite{ravipati2007toward} establishes the code context connection -- it locates the source file and line of code of the sample.
On the top, Dyninst provides information on the executed instruction and the number of bytes read from memory.

Finally, Mitos API exposes a \texttt{sample\_handler} function.
Upon sample collection, users can record additional information, such as system state, program execution-relevant, or even domain-specific information.

\subsubsection{Precision and Overhead}
Many studies have evaluated PEBS precision~\cite{yi2020precision} and overhead~\cite{akiyama2017quantitative} (or both~\cite{nonell2018applicability}), concluding that both are generally sufficient for performance analysis.
However, issues such as cache pollution~\cite{akiyama2017quantitative} or shadowing~\cite{yi2020precision} can degrade data quality.
Finally, the overhead can be very low but is highly dependent on the sampling rate.

\subsection{Output Information}

Samples form the core of the collected information.
Each PEBS sample contains the CPU (=\textbf{core ID}) and the \textbf{data source}\footnote{from which physical memory element was the piece of data loaded -- L1/L2/L3 caches, main memory, LFB (line fill buffer).}, allowing their attribution to HW resources, and also \textbf{load latency} in cycles.
Next, the \textbf{load address} is provided, which is the key for correlating the samples with the MPI buffers.
For CXL modeling, we additionally collect \textbf{timestamps} and \textbf{MPI process number} (MPI rank in relation to \texttt{MPI\_COMM\_WORLD}) for each sample via the sample handler. 
This is an extension beyond core Mitos functionality, enabling MPI context attribution.

Mitos has a predefined output structure.
Apart from the samples (CSV), it contains hwloc XML topology output, and the source code, which is intended for the MemAxes source code attribution view, the original tool on top of Mitos.

\subsection{Extensions for parallel execution}

Mitos was originally designed for sequential single-thread sampling, which has limited applicability in modern HPC.
Parallel sampling was possible but required manual API calls for each thread and extensive source code modifications -- an error-prone and cumbersome process.

\begin{figure}[t]
    \centering
    \includegraphics[width=0.42\textwidth]{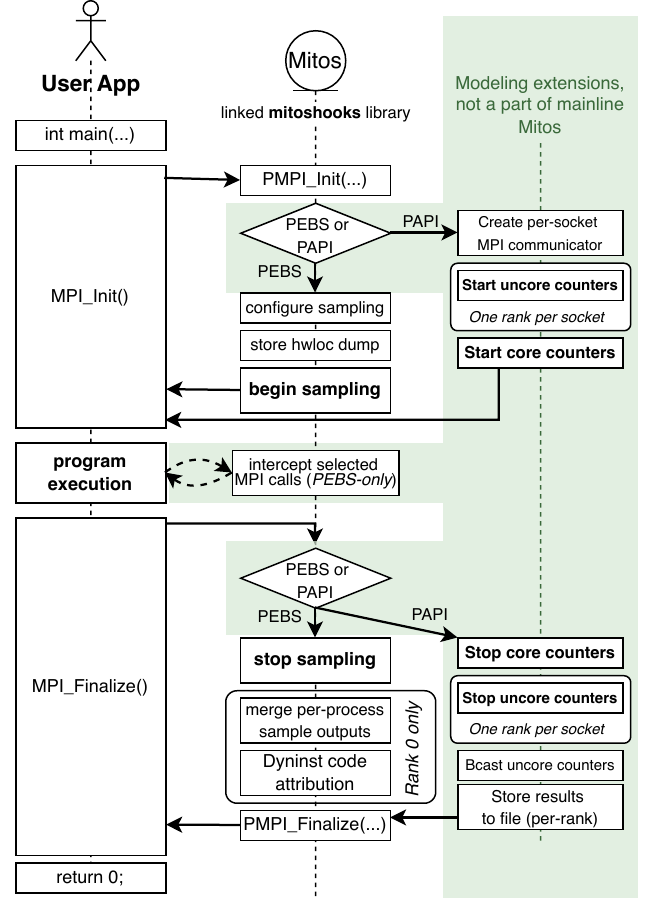}
    \caption{Workflow of Mitoshooks sampling MPI applications. The additions specifically for modeling (beyond Mitos performance analysis focus) are on the green background.}
    \label{fig:mitos-mpi-workflow}
\end{figure}

% \begin{figure}[t]
%     \centering
%     \includegraphics[width=1.02\textwidth]{fig/mitos-mpi-workflow_horizontal.drawio.pdf}
%     \caption{Workflow of Mitoshooks sampling MPI applications. The additions specifically for modeling (beyond Mitos performance analysis focus) are on the green background.}
%     \label{fig:mitos-mpi-workflow}
% \end{figure}

To better facilitate this process and significantly improve usability for parallel applications, \textbf{we extend Mitos with the capability to trace parallel (MPI and OpenMP) applications}.
We use the PMPI interface~\cite{mpi41} for MPI and the OMPT API~\cite{eichenberger2013ompt} for OpenMP programs.
The main idea behind the approach is to have an entry point to each MPI process or OpenMP thread, where Mitos sampling is configured and started, and another entry point at the end to stop the sampling and perform post-processing.
We introduce the \textbf{mitoshooks library}, which intercepts the \texttt{MPI\_Init} and \texttt{MPI\_Init\_thread} calls during initiation, and \texttt{MPI\_Finalize} during termination.
The high-level workflow is presented in Fig.~\ref{fig:mitos-mpi-workflow}.
Analogously, in OMPT, mitoshooks uses the \texttt{ompt\_initialize} callback for runtime initialization and \texttt{ompt\_finalize} for OpenMP thread termination -- and the internal logic follows the same principles.

\subsubsection{Initialization Phase}

The initialization phase resembles the original sequential Mitos execution, however, additional logic to manage the parallel execution is added.
First, \texttt{PMPI\_Init} initializes MPI to allow broadcasting of a timestamp -- an identifier of the measurement.
Next, the entrypoint virtual address is stored in a run-specific temporary file.
Then, the sampling period/frequency and the latency threshold are configured\footnote{Configured by the API or environment variables \texttt{MITOS\_LATENCY\_THRESHOLD}, \texttt{MITOS\_SAMPLING\_FREQUENCY}, \texttt{MITOS\_SAMPLING\_PERIOD}.}, the sample handler function is set, and the output structure with hwloc topology XML is created.
Finally, \texttt{Mitos\_begin\_sampler} starts the sampling and \texttt{MPI\_Init} returns to the user application.

\subsubsection{Finalizing the Sampling}

Upon the \texttt{MPI\_Finalize} call, the sampling is stopped (\texttt{Mitos\_end\_sampler}), and the post-processing begins.
First, each process concludes the measurement, stores all buffered information, and waits for other processes (\texttt{MPI\_Barrier}).
From this point, only Process 0 works on the consolidation.
It merges all process-specific outputs into a final output structure and starts the code attribution: Dyninst maps each sample IP to a specific instruction, source code line, and file\footnote{The code must be compiled with debug output (\texttt{-g}) to obtain the source code attribution context.}.

\smallskip

Since every MPI process runs its own Mitos instance, this approach is scalable and will primarily be limited by the amount of data collected and stored.
To alleviate potential bottlenecks on extremely large jobs, the sampling rate can be reduced and Mitos does not have to run on all processes.

All features mentioned in this Section \textbf{extend the overall functionality of Mitos as a performance analysis tool}, to be used together with MemAxes (white background in Fig.~\ref{fig:mitos-mpi-workflow}).
In the remainder of this Section, updates specific to this work's key topic -- performance modeling for communication improvement -- are described (in green on Fig.~\ref{fig:mitos-mpi-workflow}).

\subsection{MPI Tracing}

In addition to PEBS samples, we trace actual MPI data transfers.
We use the already-existing mitoshooks structures and add interception of more MPI calls.
For all MPI receive calls (of all types), mitoshooks records the MPI rank and tag, source rank, MPI communicator, MPI call identifier\footnote{address of the MPI call in the user application binary}, timestamps, the buffer address range, and possibly MPI requests\footnote{Other traces, such as MPI\_Wait, Init, Finalize, are collected to provide more details.}.
The buffer address range and timestamps are especially important for pairing with PEBS samples.

\subsection{PAPI Counters Collection}

MPI communication buffers -- the target of our analysis -- often exhibit data access patterns that differ significantly from the dominant application patterns.
MPI buffers tend to be used less frequently than internal data objects and rather for data exchance than internal computation.
The data access overhead part of our model consists of two steps:
First, the user collects samples through Mitos as described above.
Second, our workflow characterizes the application as a whole (cf. Sec.~\ref{sec:categorization}) to get an understanding of the overall behavior of the memory subsystem.
We add another branch to Mitos to collect PMU counters through PAPI (on the right side in Fig.~\ref{fig:mitos-mpi-workflow}), which is activated through the \texttt{MITOS\_MEASURE\_PAPI} environment variable\footnote{\texttt{MITOS\_MEASURE\_PAPI} takes name/path of the Mitos sampling output as a value.}.

We use PAPI to collect both core-level and uncore-level counters.
The uncore counters must be set up once per socket, core counters run on each process.
Therefore, mitoshooks first needs to select a leader for each socket.
First, it creates a new per-node MPI communicator.
Next, we use sys-sage~\cite{vanecek2024sys,sys-sage-poster} to separate the node into sockets by core ID\footnote{sys-sage uses the hwloc topology output created in the previous sampling run.}.
The socket ID is used to derive another MPI communicator, on which Process 0 is in charge of the uncore measurements.
Once these steps are completed, we start the time measurement and the PAPI counters.

From the core counters, the model requires the total number of LDs, L1 cache load misses, total CPU cycles, and L3 cache load misses\footnote{\texttt{PAPI\_LD\_INS}, \texttt{PAPI\_L1\_LDM}, \texttt{PAPI\_TOT\_CYC} \texttt{PAPI\_L3\_LDM}}.
The uncore counters are architecture-specific; we look for the number of loads to each memory controller on the chip.  %Intel Xeon Scalable "Cascade Lake" Gold 6240R
In our case, we set up 3 events for 3 memory controllers (IMCs), namely \\
{\footnotesize \texttt{skx\_unc\_imc[0-2/3-5]::UNC\_M\_CAS\_COUNT:RD:cpu=[cpuID]}}, on each socket.\footnote{On other architectures, we might need to recompile mitoshooks with the correct event name/number of IMCs.} 

On \texttt{MPI\_Finalize}, mitoshooks stops the timer and counter collection.
After summing up the partial per-IMC throughputs, the sum is broadcasted to the other ranks on the socket, and each MPI rank stores all counter values in the sampling output folder.

%%%%%%%%%%%%%%%%%%%%%%%%%%%%%%%%%%%%%%%%%%%%%%%%%%%%%%%%%%%%%%%%%%%%%%%%%%%%%%
\section{MPI and CXL communication model}\label{sec:model}

With all relevant information collected by mitoshooks, the next step is to process the raw data and generate performance predictions.
For this we develop a novel model combining memory and communication behavior, which evaluates both:
\begin{enumerate}
    \item The actual (baseline MPI) and predicted (CXL) data transfer overhead
    \item The data load overhead from memory segments with different performance (caches, main memory, or CXL)
\end{enumerate}
We compute the combined overhead -- data transfer plus data access -- for both MPI and CXL scenarios, separately for each MPI receive call in the source code.
The final result for each MPI call is the difference between the MPI and CXL options.
Specific HW capabilities and model thresholds are defined via model parameters, further described in this Section.

\subsection{Data Transfer Overhead}

Data transfer overhead is the simpler component of our model.
We apply the Hockney model~\cite{hockney1994communication} to estimate MPI data transfer overhead.
It has the following form: 

\begin{equation}
T_{\substack{Transfer \\MPI}} = \sum_{MPI\_Trace} ( MPI\_LAT + bufsz / MPI\_BW )
\end{equation}
where \( bufsz \) is the MPI buffer size of each trace and $MPI\_LAT$ and $MPI\_BW$ are MPI bandwidth and latency model parameters.

Since the data transfer overhead computation is isolated from the Data Access Overhead computation, the model can easily be replaced by, for instance, one of the LogP-family models~\cite{hoefler2010loggopsim}.

In CXL-based communication, explicit data transfer is absent; however, synchronization between sender and receiver is required.
We assume an atomic locking mechanism for synchronization: the sender signals "ready-to-read", and the receiver signals "ready-to-write" to free the buffer for subsequent transfers.
This synchronization is the only communication overhead, which we model as \texttt{CXL\_ATOMIC\_LAT}\footnote{We observed ca. \SI{150}{\nano\second} longer atomic load times compared to regular loads from the same memory in our benchmarks; we set this parameter accordingly.} for each notification.
\begin{equation}
T_{\substack{Transfer \\CXL}} = \sum_{MPI\_Trace} (2* CXL\_ATOMIC\_LAT)
\end{equation}

\subsection{Application Characterization}\label{sec:categorization}

Our \textbf{data access overhead} computation analyzes the impact of collected load (LD) samples on MPI buffers based on the workload type.
For instance, a cache miss sample in a high-throughput section of code, such as streaming access to data, will be strongly impacted by being allocated on a different physical memory (i.e., CXL) with lower bandwidth.
Contrary to that, workloads with good cache usage do not stress the memory bandwidth, and hence a CXL LD will be less impacted.
More generally, there are slight differences in the predictions for a sample, depending on what the application characteristic is with respect to access to data. 
We classify workloads into five categories:
\begin{itemize}
    \item Memory bandwidth limited
    \item Memory latency limited
    \item Cache bandwidth limited
    \item Cache latency limited
    \item Compute (does not strongly fall into any of the above; can also be I/O, communication, network, etc.)
\end{itemize}

Each category is assigned a weight in the range [0,1], normalized so that the sum of all equals 1.
This characterization assesses the overall memory subsystem stress, not just the MPI buffers, so we obtain the necessary information from PMU counters (PAPI).
In this study, we characterize the application over the entire runtime; alternatively, querying these counters periodically would provide a more fine-grained view, capturing different application stages.

\subsubsection{Computing the Weights}
Each category's weight is determined by a specific metric, based primarily on PAPI counters, and so-called lower and upper thresholds, which are model parameters.
The weight $W$ is 0 if the value is below the lower and 1 if above upper threshold.
The weight increases quadratically between thresholds -- remaining low near the lower bound and rising sharply toward the upper bound -- allowing wide threshold ranges:
\begin{equation}
	W_{Characteristic} = 
	\begin{cases}
		0 &\text{if } val \leq THR\_L \\
		1 &\text{if } val \geq THR\_U \\
		(\frac{val - THR\_L}{THR\_U - THR\_L})^2 & \text{otherwise}
	\end{cases}
\end{equation}
where $val$ is the specific metric for each characteristic, $THR\_L$ and $THR\_U$ are the lower and upper thresholds.

\textit{Memory bandwidth} (MBW) metric is the average memory throughput on-socket, and the thresholds are percentages of the peak memory bandwidth.

\textit{Memory latency} (MLAT) metric is the percentage of L3 cache LD misses among all LD instructions, and the thresholds define the minimum and maximum to consider.
As the MLAT metric may be high for the MBW category as well, the resulting MLAT weight is $W_{MLAT}\!-\!W_{MBW} \text{ (but never \textless 0)}$.

The \textit{cache bandwidth} (CBW) and \textit{cache latency} (CLAT) weights are computed conceptually similarly to the memory counterparts.
For CBW, the metric is the cache throughput (\unit[per-mode = symbol]{\byte\per\second}), and the thresholds are percentages of the cache bandwidth.
This metric is measured for both L1 and L2, the resulting $W_{CBW}$ is the maximum of both.
CLAT only considers L2 cache, as the L1 latency is negligible for an application performance profile.
The value is the fraction of all LDs that reach the L2 cache, and the thresholds define the minimum and maximum to consider.
Analog to MLAT, CLAT deducts all above-mentioned weights, as this metric will be high for MBW, MLAT, CLAT as well:
\begin{equation}
\scalebox{0.85}{
$\displaystyle
W_{CLAT} = W_{CLAT} - (W_{MBW}\! +\! W_{MLAT}\! +\! W_{CBW}) \qquad\text{(but never \textless 0)}
$
}
\end{equation}

\subsubsection{First and Subsequent Loads}

After the sender writes data, the receiver loads it from CXL memory.\footnote{Prefetching may bring the data to the caches but still originates in the memory.}
The subsequent loads will be from cache, or CXL if evicted from caches.
This results in two sets of weights: 
\begin{enumerate}
    \item For the first load (after the sender writes new data), involving a CXL read, we consider MBW, MLAT, and Compute categories, as cache characteristics are less relevant.
    \item For the subsequent accesses, which may or may not involve CXL, we already consider all five categories.
\end{enumerate}
As mentioned above, the weights for MBW, MLAT, Compute are normalized to sum up to 1 in Case (1) and in Case (2), weights for MBW, MLAT, CBW, CLAT and Compute are normalized to sum up to 1 as well.\footnote{If all weights sum up to less than 1, the remainder is assigned to Compute, up until the \texttt{COMPUTE\_MAX\_WEIGHT} parameter. 
Above this value, the remainder to 1 is split equally by all other categories. In the case when weights sum up to more than 1, each weight is divided by the sum.}
There are no means to determine which group a sample belongs to, we only know how many accesses on average to each element/buffer.
So, $1/n$-th of each sample is considered a first access, and $(n-1)/n$-ths are considered subsequent access.

\subsection{Data Access Overhead}\label{sec:data-access-overhead}

Data access overhead is computed from each sample's latency and data source, scaled by the sampling period to estimate the total overhead of all load operations.
To convert sample load latency into overall program overhead, we account for parallelization: 
In perfectly latency-limited codes, loads are serialized, so no parallelization occurs.\footnote{On modern CPUs with out-of-order execution, computation can still take place while waiting for a load.}
In perfectly bandwidth-limited codes, multiple loads are issued concurrently, limited primarily by the load queue size (typically tens to hundreds on modern CPUs). 
Hence, the sample latency contributes to the overall overhead only by a fraction.
This concept is illustrated on Fig.~\ref{fig:bw-lat-limited}.
Although no code is purely bandwidth- or latency-limited, parallelism varies. 
Our model accounts for this using two factors: \texttt{LPF\_BW} for bandwidth-limited and \texttt{LPF\_LAT} for latency-limited workloads.

\begin{figure}[t]
    \centering
    \includegraphics[width=0.45\textwidth]{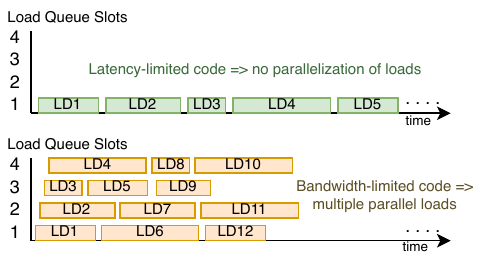}
    \caption{Illustrative examples of load parallelism under latency-limited and bandwidth-limited conditions.}
    \label{fig:bw-lat-limited}
\end{figure}

In some cases, MPI receive buffers are subsets of larger internal arrays, enabling direct placement of received data. 
This is not feasible with CXL, as data resides in a separate physical memory region for communication purposes.
Therefore, an unpack operation -- copying data from CXL to the target internal buffer -- is required.
The expensive CXL access takes place only once during unpack; the other loads target DDR directly.
Our model allows users to optionally model this scenario -- through modeling only a fraction of each sample\footnote{The fraction normalizes the number of collected samples to one LD per element.} as CXL access and the remainder as DDR accesses.

In the \textbf{MPI scenario}, the model sums sample latencies ($lat$) for each MPI call and normalizes them by the sampling rate ($rate$) and the parallelization factor ($lpf$):

\begin{equation}
T_{\substack{Access \\MPI}} = (\sum_{LD\_Trace} lat) * rate / lpf
\end{equation}

\begin{figure}[t] 
    \centering
    \includegraphics[width=0.47\textwidth]{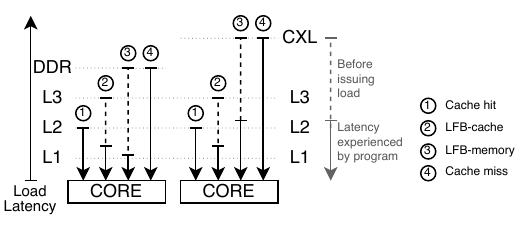}
    \caption{Comparison of DDR and CXL performance across different data sources (cache hits, LFB, memory).}
    \label{fig:ld-scenarios}
\end{figure}

The \textbf{CXL scenario} models expected behavior using different memory elements, reflecting their impact on performance.
Each of the categories mentioned in Sec.~\ref{sec:categorization} uses a slightly different formula for calculating the overhead, reflecting the different usage pattern of the memory elements.
At a high level, we distinguish samples impacted by switching to CXL from those unaffected.
A good indicator is the \textit{data source} of the sample.
Fig.~\ref{fig:ld-scenarios} presents the changes in the load times resulting from different physical data location (\textit{data source}).
Cache hits exhibit similar performance in both DDR (MPI) and CXL scenarios, unless the piece of data was prefetched.
Some of the cache hits may be caused by prefetching the cache line from the fast DDR before it is needed.
We cannot assume that prefetching from the slower CXL will complete always in time as well.
Additional uncertainty arises from LFB (Line Fill Buffer) samples.
It practically says that the cache line was being fetched in the moment of taking the sample, however, it does not say from where -- from a higher level cache or memory.
An \textit{LFB} originating in cache will have the same overhead, while an LFB from the DDR and CXL will perform differently.
Finally, the overhead of DDR and CXL accesses will clearly differ as well.

To target the prefetching implications, the model distinguishes demand and prefetch cache hits.
Prefetch hits in DDR may become cache misses or LFB samples under CXL, particularly under high bandwidth pressure.
Therefore, we model different overhead for demand and prefetch cache hits\footnote{There is no instrument to tell which sample is a prefetch or demand hit. The model assumes that one LD in each cache line is \textbf{not} a demand hit. A fraction of each sample is considered demand hit and the remainder (0 if the demand hit percentage exceeds the overall hit percentage) a prefetch hit.}.

As a result, the total data access overhead of the CXL scenario is a sum of all the partial sample overheads (as in the MPI version), however, the samples are grouped into different brackets by the data source, with each bracket having a different formula for the overhead.
These formulas differ for each category defined in Sec.~\ref{sec:categorization}.

\paragraph{\textbf{Memory Latency}}
Under low bandwidth pressure, we optimistically assume timely prefetching. 
Conversely, we pessimistically assume LFB traces incur longer load times.
It is computed as follows:

\begin{equation}
\scalebox{0.85}{
$\displaystyle
    \begin{split}
T_{\substack{Access \\CXL}} &= (\sum_{\substack{LD\_Trace \\ Cache Hit}} lat \\
&+ \sum_{\substack{LD\_Trace \\ LFB }} (lat\! +\! CXL\_LAT\! -\! MEM\_LAT) \\
&+ \sum_{\substack{LD\_Trace \\ Cache Miss}} CXL\_LAT ) * rate / LPF\_LAT
    \end{split}$
    }
\end{equation}
where $CXL\_LAT$ and $MEM\_LAT$ are model latency parameters for CXL and main memory (DDR).

\paragraph{\textbf{Memory Bandwidth}}
The MBW category distinguishes demand and prefetch hits.
Due to the high memory pressure, prefetch hits may hit the LFB or miss the cache entirely (pessimistic assumption). 
Next, as shown in~\cite{radulovic2019profet}, programs experience larger latencies when utilizing higher bandwidth.
To account for congestion in the memory subsystem, the model allows the latency to be higher than the \texttt{CXL\_LAT} parameter.
This scenario is modeled as follows:

\begin{equation}
\scalebox{0.85}{
$\displaystyle
    \begin{split}
T_{\substack{Access \\CXL}} &= (\sum_{\substack{LD\_Trace \\ Demand Cache Hit}} lat  \\
&+ \sum_{\substack{LD\_Trace \\ Prefetch Cache Hit }} (lat\! +\! CXL\_LAT\! -\! MEM\_LAT)  \\
&+ \sum_{\substack{LD\_Trace \\ LFB }}  \max \{CXL\_LAT , (lat\! +\! CXL\_LAT\! -\! MEM\_LAT) \}  \\
&+ \sum_{\substack{LD\_Trace \\ Cache Miss}} \max \{CXL\_LAT , (lat\! +\! CXL\_LAT\! -\! MEM\_LAT) \}  )  \\
&* rate / LPF\_BW
    \end{split}$
    \raisetag{15pt}
    }
\end{equation}

\paragraph{\textbf{Cache Bandwidth}}
CBW resembles the MBW category, but differs in one aspect:
the low memory usage indicates that the traffic takes place mainly in the cache, so the model (optimistically) assumes that LFB samples originate in the cache.

\begin{equation}
\scalebox{0.85}{
$\displaystyle
    \begin{split}
T_{\substack{Access \\CXL}} = &(\sum_{\substack{LD\_Trace \\ Demand Cache Hit}} lat \\
&+ \sum_{\substack{LD\_Trace \\ Prefetch Cache Hit}}  (lat\! +\! CXL\_LAT\! -\! MEM\_LAT)  \\
&+ \sum_{\substack{LD\_Trace \\ LFB }} lat \\
&+ \sum_{\substack{LD\_Trace \\ Cache Miss}} \max \{CXL\_LAT , (lat\! +\! CXL\_LAT\! -\! MEM\_LAT) \}  )  \\
&* rate / LPF\_BW
    \end{split}$
    }
    \raisetag{15pt}
\end{equation}

\paragraph{\textbf{Cache Latency}}
This category shares the structure and reasoning of MLAT.
As with CBW, the algorithm uses mainly cache, which indicates good temporal and spatial locality of the code.
That is a precondition for efficient prefetching, so we (optimistically) assume prefetching from memory not being an issue, given the low memory throughput.
Also, the LFB samples are likely originating in the cache rather than memory, hence, no CXL penalty there either:

\begin{equation}
\scalebox{0.85}{
$\displaystyle
    \begin{split}
T_{\substack{Access \\CXL}} = &(\sum_{\substack{LD\_Trace \\ Cache Hit}} lat \\
&+ \sum_{\substack{LD\_Trace \\ LFB }} lat \\
&+ \sum_{\substack{LD\_Trace \\ Cache Miss}} CXL\_LAT ) * rate / LPF\_LAT
    \end{split}$
    }
\end{equation}

\paragraph{\textbf{Compute}}
As the Compute category is not limited by the data accesses, we assume good parallelism, so we use \texttt{LPF\_BW}.
Given the low throughput, we assume prefetchers work well (as in MLAT and CLAT scenarios) and finally, for the LFB traces, we average the MLAT and CLAT approaches -- assuming half of the LFB traces comes from the cache and other half from the DDR:

\begin{equation}
\scalebox{0.85}{
$\displaystyle
    \begin{split}
T_{\substack{Access \\CXL}} = &(\sum_{\substack{LD\_Trace \\ Cache Hit}} lat \\
&+ \sum_{\substack{LD\_Trace \\ LFB }} (lat + \frac{CXL\_LAT - MEM\_LAT}{2}) \\
&+ \sum_{\substack{LD\_Trace \\ Cache Miss}} CXL\_LAT ) * rate / LPF\_BW
    \end{split}$
}
 \raisetag{15pt}
\end{equation}

%%%%%%%%%%%%%%%%%%%%%%%%%%%%%%%%%%%%%%%%%%%%%%%%%%%%%%%%%%%%%%%%%%%%%%%%%%%%%%
\section{Use Cases}

We showcase the functionality and accuracy of our analysis and our model on two use cases -- Heat Transfer on 2D Plane and the HPCG benchmark. 
Since the target CXL.mem hosts have yet to enter the market, we evaluate the use cases on a single node setup with heterogeneous memory to cross-validate the quality of the model prediction against a reference shared memory implementation. 
On the top, we show-case the intended setup through presenting the model prediction in a multi-node scenario assuming a future CXL.mem implementation using expected system characteristics for such architectures.

\subsection{Experimental Setup}

\begin{figure}[tb]
  \centering
  \begin{subfigure}{0.49\columnwidth}
    \includegraphics[width=\linewidth]{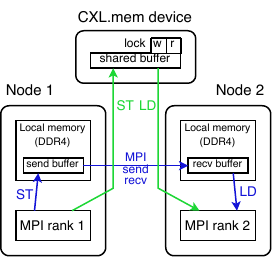}
    \caption{Scenario with pooled CXL.mem}
    \label{fig:sub1}
  \end{subfigure}
  \hspace{0.005\columnwidth}
  \begin{subfigure}{0.49\columnwidth}
    \includegraphics[width=\linewidth]{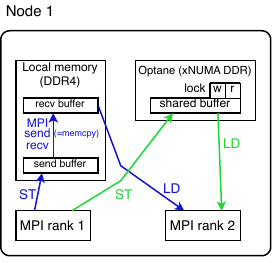}
    \caption{One-node validation scenario}
    \label{fig:sub2}
  \end{subfigure}
  \caption{Data allocation and communication paths}
  \label{fig:comm}
\end{figure}

We validate the model results on a single-node setup using a reference implementation.
Our test system consists of two Intel Xeon Gold 6240R sockets (Cascade Lake), each with 24 cores/48 threads at 2.40 GHz, 36 MB L3 cache, and 200 GB DDR4 per NUMA region, complemented by 2×744 GB Optane Persistent Memory.
Our test system consists of two Intel Xeon Gold 6240R CPU sockets (Cascade Lake), each with 24 cores/48 threads (@2.40GHz), \SI{36}{\mega\byte} L3 cache, and \SI{200}{\giga\byte} DDR4 in 1 NUMA region, complemented by 2x\SI{744}{\giga\byte} Optane Persistent Memory.
DDR and Optane form an on-node heterogeneous memory setup with two different latencies, used to emulate the effect of higher-latency CXL.mem.
We compiled all codes with gcc 11.4.0 (-O3) and run with OpenMPI 5.0.6.

The multi-node runs (Sec.~\ref{sec:multinode}) use four Intel Skylake Xeon Platinum 8174 nodes, each with 2 sockets (24 cores/48 threads @3.10 GHz), \SI{36}{\mega\byte} L3 cache, and \SI{96}{\giga\byte} DDR4. 
The code was compiled and run with gcc 11.2.0 (-O3) and OpenMPI 4.1.2\footnote{We used the native MPI platform on both systems, hence the used versions differ.}.

\subsection{Setting Model Parameters}
We use multiple benchmarks to derive hardware and middleware parameters presensed in Sec.~\ref{sec:model}: 
\begin{enumerate}
  \item We use OSU Micro-Benchmarks~\cite{osu_mb} (7.5.1), and its \texttt{osu\_bw} and \texttt{osu\_latency} to retrieve MPI-related parameters, namely \textit{MPI\_BW}: \SI{9.444}{\giga\byte\per\second} and \SI{4.090}{\giga\byte\per\second} for on-NUMA (Sec.~\ref{sec:hpcg}) and cross-NUMA transfer (Sec.~\ref{sec:stencil}), respectively; and \textit{MPI\_LAT}: \SI{320}{\nano\second} and \SI{650}{\nano\second}. 
  \item We use \textit{likwid-bench}~\cite{treibig2012likwid} (5.4.1) to benchmark L2 cache (\SI{52}{\giga\byte\per\second}) and main memory (\SI{73}{\giga\byte\per\second}) bandwidth.
  \item We wrote custom microbenchmarks to determine the main memory (DDR) latency, the CXL latency and the atomic operation latency on CXL. These p-chase-like benchmarks are available as a part of our toolchain as well. Measured DDR latency is \SI{86}{\nano\second}. CXL latency is set to \SI{86}{\nano\second} when mimicked by on-NUMA DDR, \SI{154}{\nano\second} by cross-NUMA DDR, and \SI{417}{\nano\second} when mimicked by Optane. The CXL atomic latency is \SI{653}{\nano\second} for Optane, and \SI{191}{\nano\second} and \SI{210}{\nano\second} on on-NUMA and cross-NUMA DDR, respectively.
\end{enumerate}

The remaining parameters, particularly model thresholds, cannot be benchmarked and are set heuristically.
We use (in the form "lower--upper") \SI{0.03}-\SI{0.33} for MBW, \SI{0.01}-\SI{0.2} for MLAT, \SI{0.1}-\SI{0.75} for CBW, and \SI{0.05}-\SI{0.5} for CLAT.
\texttt{LPF\_LAT} is set to \SI{1.5}, \texttt{LPF\_BW} is \SI{3}, and \texttt{COMPUTE\_MAX\_WEIGHT} is \SI{0.5}.

\subsection{Heat Transfer on 2D Plane}\label{sec:stencil}

\begin{figure*}[tb]
  \centering
  \begin{subfigure}[t]{0.499\textwidth}
    \centering
    \includegraphics[width=\linewidth]{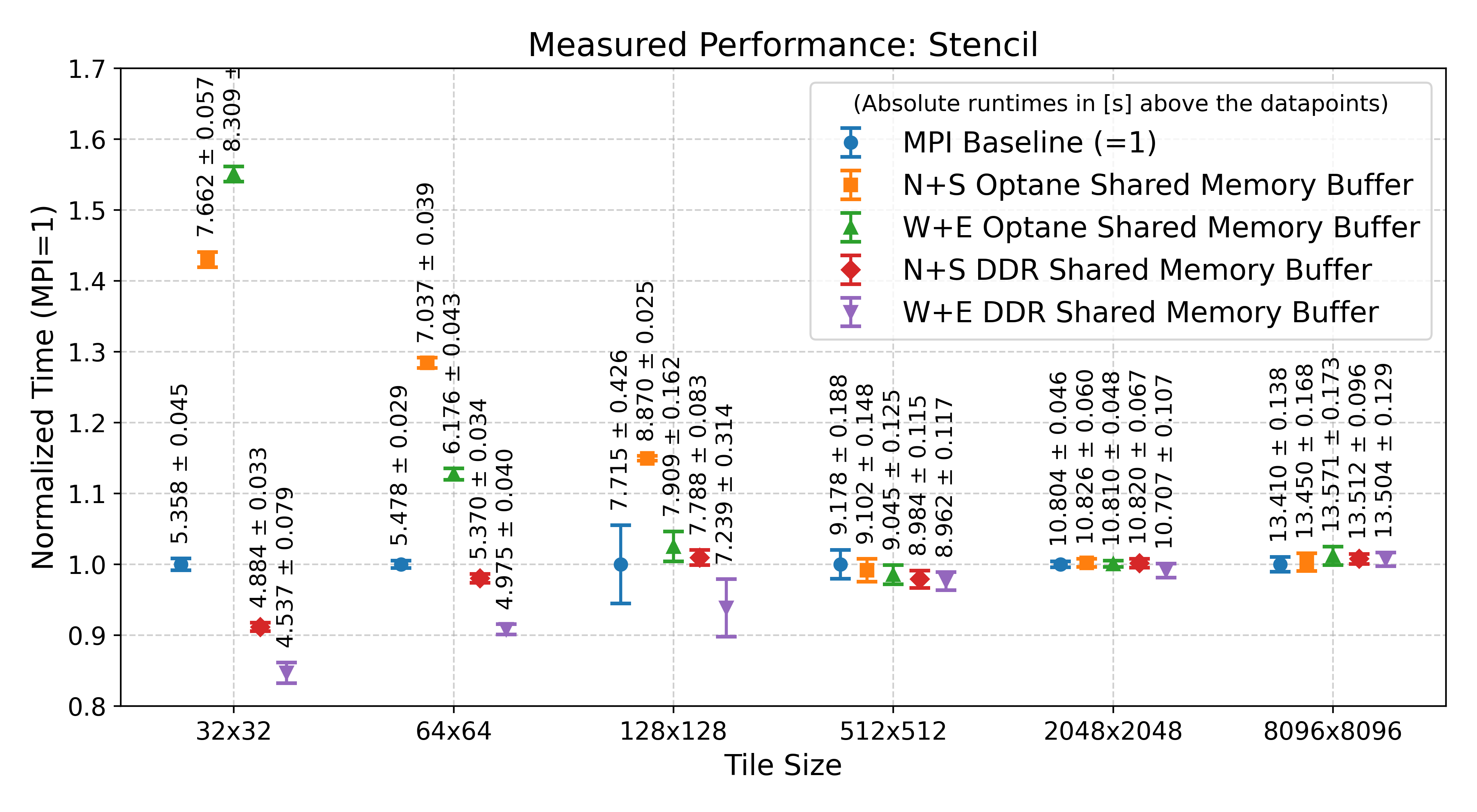}
    \caption{Reference implementation performance on one node.}
    \label{fig:stencil-real}
  \end{subfigure}\hfill
  \begin{subfigure}[t]{0.499\textwidth}
    \centering
    \includegraphics[width=\linewidth]{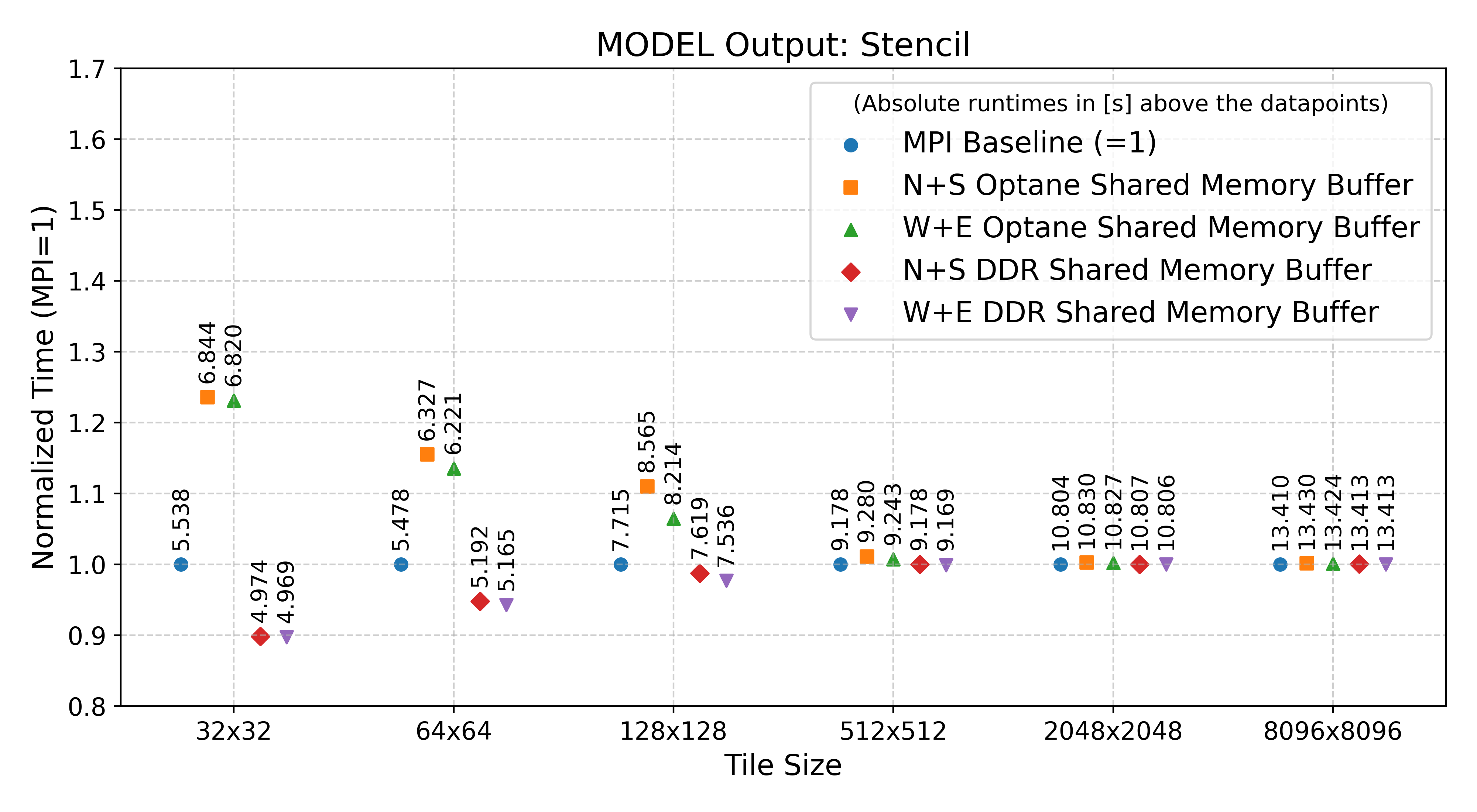}
    \caption{Model prediction (should reflect the reference implementation).}
    \label{fig:stencil-model}
  \end{subfigure}
  \caption{Comparison of model predictions and reference implementation for 2D stencil with N+S and W+E halo exchanges using shared memory.}
  \label{fig:stencil-perf}
\end{figure*}

The first use-case is a heat transfer simulation on a 2D plane, approximating the solution of a partial differential equation using a 5-point stencil code.
The simplicity of this stencil code facilitates clear analysis of application behavior and model predictions.
The code uses MPI for parallelization, dividing the 2D plane into smaller tiles, with each MPI process computing one tile.
Between time steps, MPI ranks exchange boundary regions (halo layers) to update cell values computed by neighbors.
Therefore, the cells neighboring the halo layer compute the new value partly from the internal tile cells and partly from the halo buffer.\footnote{Halos are not unpacked, they remain as contiguous memory in the receive buffers.}
The halo exchange, in form of MPI send-recv, is the target of this analysis as potential candidates to CXL offload.

\subsubsection{Model validation}

The original implementation uses MPI (\texttt{MPI\_Isend} and \texttt{MPI\_Irecv}) to communicate between the neighbors, and we implement an additional shared memory-based approach as a reference for the model predictions. 
The shared memory buffers can be allocated on both Optane and cross-NUMA DDR, in both cases simulating memory that is slower to access than on-NUMA DDR.
This setup mimics the target scenario -- a multi-node setup with a shared memory buffer on the (slower) CXL.mem pooled memory exposing the halo buffers.
The only difference between our test and the target CXL scenarios are the model parameters (such as CXL vs. Optane latency, or on-node vs. cross-node MPI transfer).
Both the test and target scenarios are sketched in Fig.~\ref{fig:comm} with data movements in blue for MPI and green for (CXL) shared memory buffer communication.

We use 16 (4x4) tiles to split the problem, allocated in a chessboard-like pattern on the two NUMA domains to ensure the data moves between sockets (not staying in lower-level caches).
Since there are 4 neighbors and 4 halos on each process (excluding the edges) -- N, S, W, E -- we make measurements replacing only the horizontal ones (N+S) and only the vertical ones (W+E), while other halos stay as MPI exchange.
Having 2 additional memory types from the receiver's perspective (Optane and off-NUMA DDR), we compare 5 scenarios:
\begin{itemize}
    \item Baseline MPI
    \item N+S Optane (and W+E MPI)
    \item W+E Optane (and N+S MPI)
    \item N+S off-NUMA DDR (and W+E MPI)
    \item W+E off-NUMA DDR (and N+S MPI)
\end{itemize}

We run all scenarios with different tile sizes (from 32x32 to 8096x8096 cells), and measure the execution time.
The reference results are shown on Fig.~\ref{fig:stencil-real}: 
The time (normalized by the reference MPI implementation time) is on the y axis, the tile sizes are on the x axis (with absolute values and stdev next to the data points).
Different scenarios are presented with different colors.

% \begin{figure}
% \centering
% \begin{minipage}{.5\textwidth}
%   \vspace{11pt}
%   \centering
%   \includegraphics[width=.8\linewidth]{fig/stencil-access.drawio.pdf}
%   \caption{Temporal locality of N and W halo accessed differs}
%   \label{fig:stencil-access}
% \end{minipage}%
% \begin{minipage}{.5\textwidth}
%   \centering
%   \includegraphics[width=.99\linewidth]{fig/plot_stencil_multinode.png}
%   \caption{Temporal locality of N and W halo accessed differs}
%   \label{fig:stencil-multinode}
% \end{minipage}
% \end{figure}

\begin{figure}[t]
    \centering
    \includegraphics[width=0.34\textwidth]{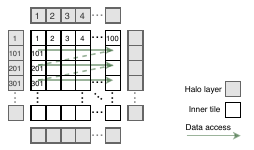}
    \caption{Differences in temporal locality between N and W halo accesses.}
    \label{fig:stencil-access}
\end{figure}

\begin{figure}[t]
  \centering
  \includegraphics[width=0.495\textwidth]{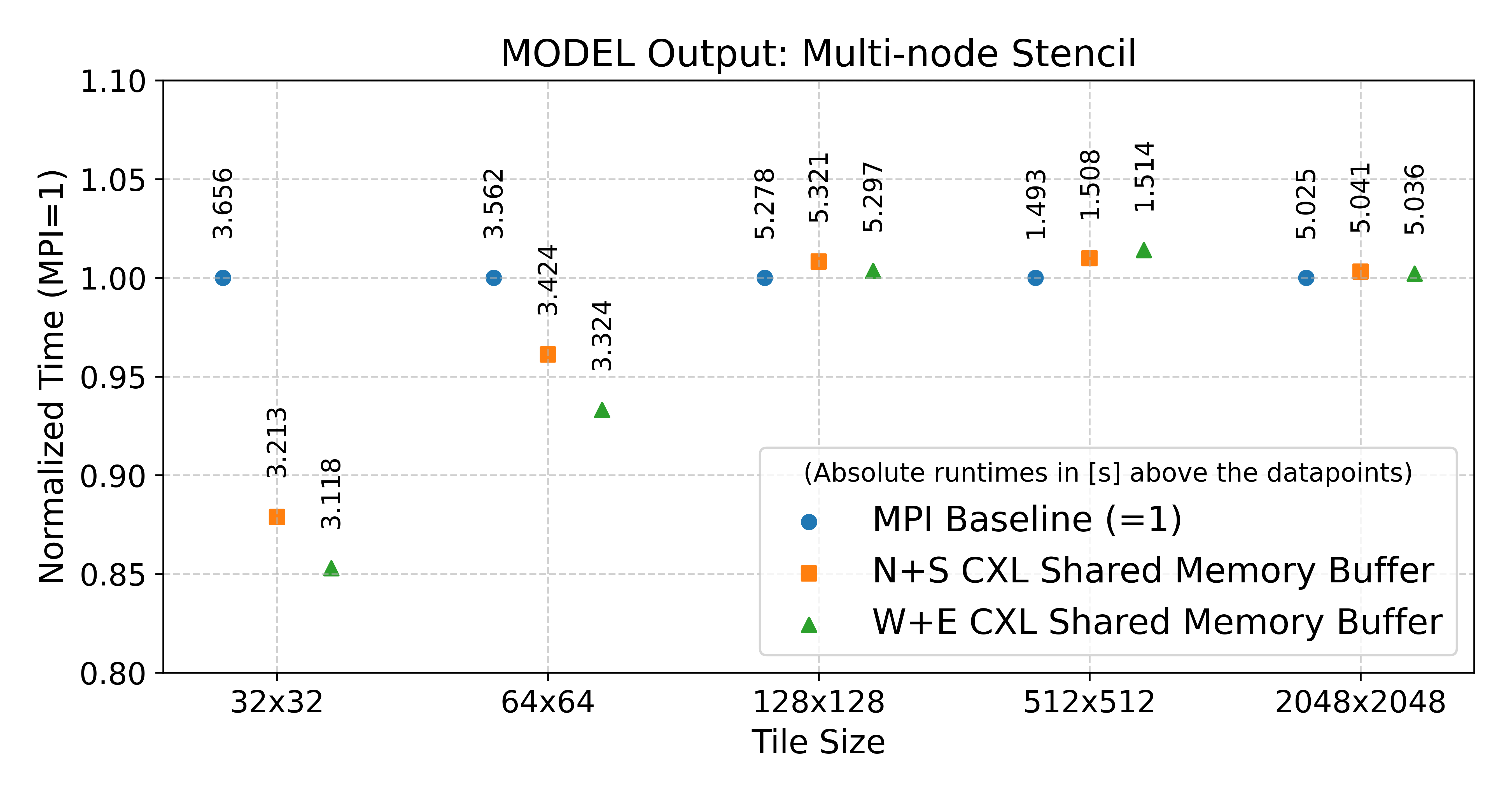}
  \caption{Predicted performance of CXL.mem-based communication on a four-node setup.}
  \label{fig:stencil-multinode}
\end{figure}

We observe several key trends:
(1) Performance impact varies significantly, from 1.22x speedup to 0.67x slowdown.
(2) Performance differences are much more significant on the smaller tile sizes and hardly noticeable on the large ones.
The computation-to-communication ratio changes (8:1 for 32x32 vs 2048:1 for 8096x8096\footnote{4x32 halos vs 32x32 cell computations on the smallest size and 4x8096 halos vs 8096x8096 cell computations on the largest time size.}), so the overall impact of the halo layers is much smaller.
(3) The Optane implementation is slower than DDR\footnote{8096x8096 is an exception, however, the difference is less than one standard deviation.} -- this is understandable due to better performance of DDR.
(4) Overall, W+E halo exchanges outperform N+S, except for a few outliers.
The halos are contiguous buffers of the same size, so data transfer performance does not differ, unlike the access pattern to these buffers, as illustrated on Fig.~\ref{fig:stencil-access}.
All halos are accessed in a streaming fashion -- conductive to prefetching.
The difference is in the temporal locality of the accesses: horizontal halo is accessed in one batch, whereas the vertical one with long breaks in between (to compute one row of the tile).
Intuitively, better temporal locality should yield superior performance; however, our results indicate the opposite.
We attribute this to the inner tile remaining in low-level caches while halo data resides in slower memory after transfer.
Although the prefetcher correctly anticipates the next cache line, horizontal halos require it sooner than it can arrive, whereas vertical halos allow more time before the next line is needed.
With larger tile size, the vertical halos may face cache evictions, hence losing the advantage, as we see in the results.

We compare the real performance values with our model predictions.
We link the mitoshooks library to the code: it generates the samples and PAPI metrics in two application runs with zero user intervention.
Then we run the model computation.
Fig.~\ref{fig:stencil-model} presents the predicted performance.
We observe all the trends as mentioned above -- 1) the projected speedup differs considerably (1.11x to 0.81x), 2) the largest differences are on smaller tile sizes, 3) the Optane implementation is always slower than DDR, and 4) the W+E buffers as shared memory perform better than N+S.
Except for the smallest tile size (32x32), predicted values closely match observed performance.
Our approach successfully achieves its goal: guiding developers on when to adopt shared memory data exchange.
They are clearly aided to prioritize W+E halos over N+S, to not use Optane, as it is too slow, and to only consider cross-NUMA DDR if their tile size is small -- so that it brings a measurable benefit.

\begin{figure}[t]
  \centering
  \includegraphics[width=0.49\textwidth]{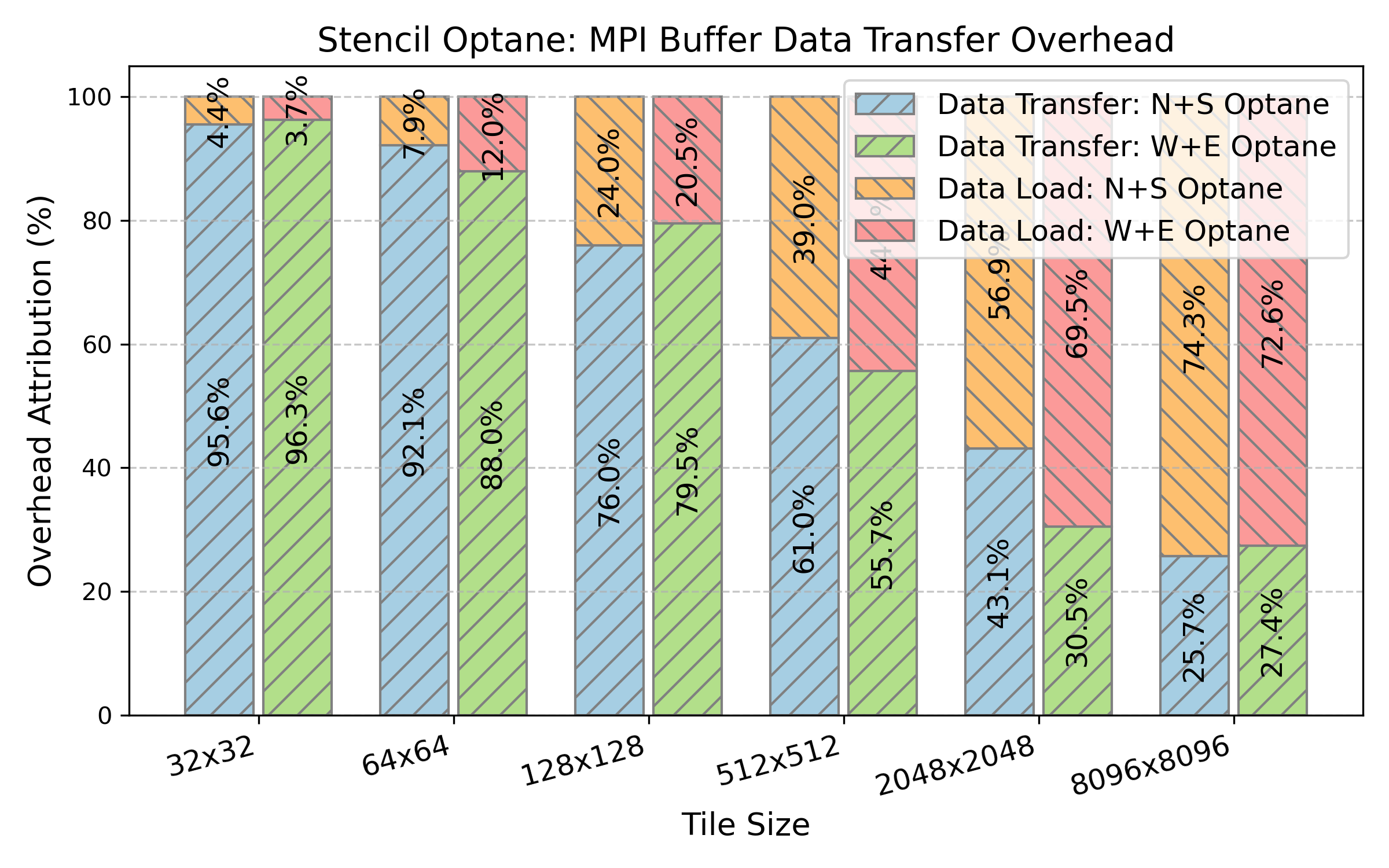}
  \caption{Breakdown of data load and transfer costs for stencil computation for different tile sizes.}
  \label{fig:stencil-breakdown}
\end{figure}

\subsubsection{Overhead Breakdown}\label{sec:stencil-overhead-breakdown}
Both data loads (from caches or memory) and transfers (MPI or CXL) contribute to overall overhead.
Fig.~\ref{fig:stencil-breakdown} breaks down the modeled Optane shared memory buffer overhead into these two categories for horizontal and vertical halos.
For small tile sizes, data transfer overhead dominates due to numerous small transfers, rendering data load overhead negligible (e.g., \SI{4}{\percent} for 32x32). 
As tile size increases, data load overhead becomes dominant, accounting for up to \SI{74}{\percent} of total overhead.
This breakdown shows that, depending on the communication and data access patterns, either factor can dominate the cost function, making both crucial for a precise prediction.

\begin{figure*}[tb]
  \centering
  \begin{subfigure}[t]{0.499\textwidth}
    \centering
    \includegraphics[width=\linewidth]{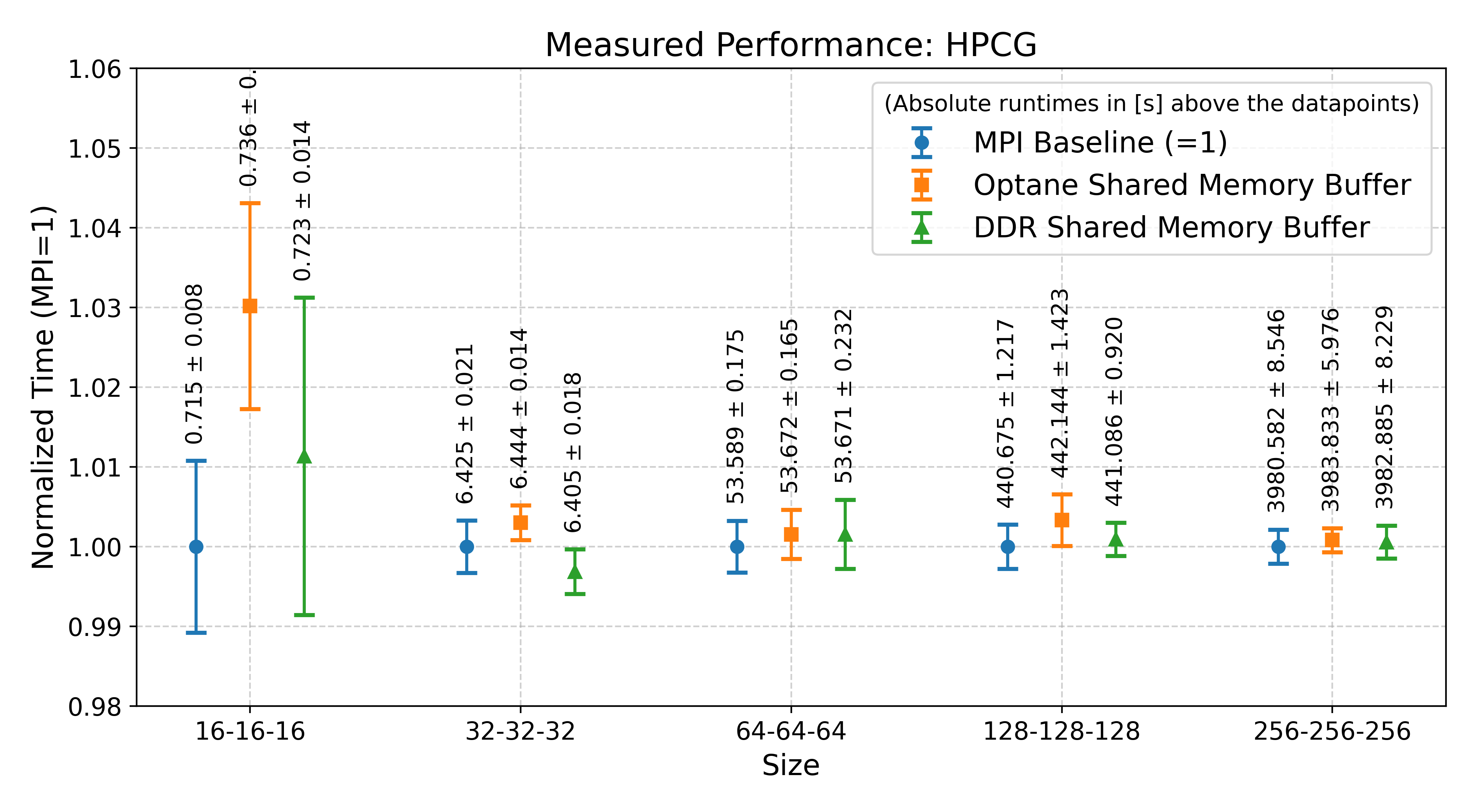}
    \caption{HPCG reference implementation performance on one node.}
    \label{fig:hpcg-real}
  \end{subfigure}\hfill
  \begin{subfigure}[t]{0.499\textwidth}
    \centering
    \includegraphics[width=\linewidth]{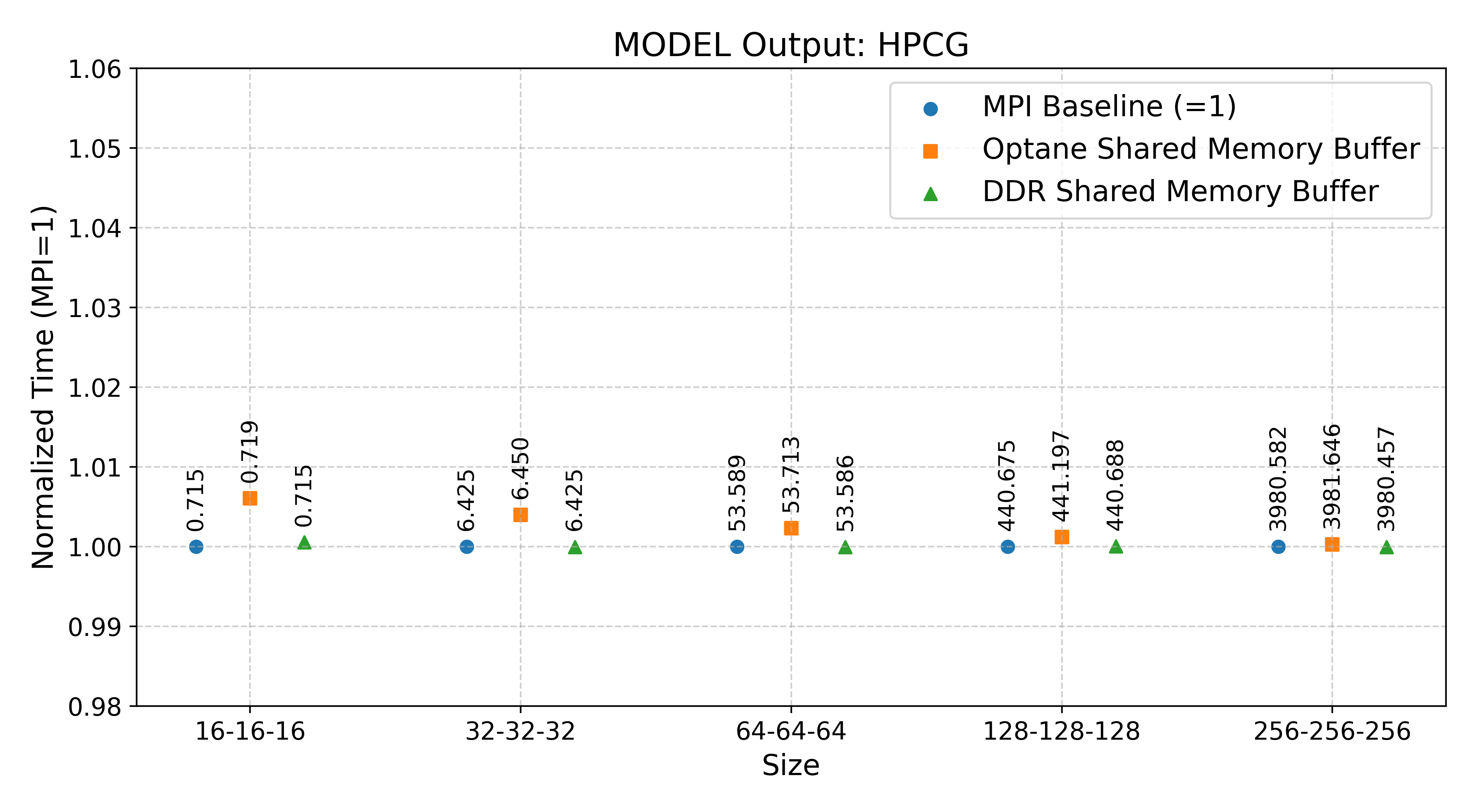}
    \caption{Model prediction (should reflect the reference implementation).}
    \label{fig:hpcg-model}
  \end{subfigure}
  \vspace{-8pt}
  \caption{Comparison of model predictions and reference implementation for HPCG benchmark using DDR and Optane shared memory.}
  \label{fig:hpcg-perf}
  \vspace{-10pt}
\end{figure*}

\subsubsection{Multi-node prediction}\label{sec:multinode}

The second evaluation scenario of this use case is a multi-node setup.
We allocate 64 (8x8) tiles across 4 nodes, always enforcing cross-node communication.
We run mitos data collection and generate model prediction for tile sizes from 32x32 to 4096x4096.
We benchmark {MPI\_LAT} as \SI{1.48}{\micro\second} and \texttt{MPI\_BW} as \SI[per-mode = symbol]{24.715}{\giga\byte\per\second}. We set \texttt{CXL\_LAT} to \SI{350}{\nano\second} (average of \SIrange{300}{400}{\nano\second} mentioned in~\cite{ahn2024mpi}) and \texttt{CXL\_ATOMIC\_LAT} to \SI{430}{\nano\second}.\footnote{The CXL-related values are just illustrational; it is yet to be verified/benchmarked on real systems when they arrive.}

As CXL.mem 3.0+ hardware is not yet available, no reference implementation exists. Therefore, only the model predictions are presented in Fig.~\ref{fig:stencil-multinode}. 
The pattern resembles the on-node DDR shared memory performance (Fig.~\ref{fig:stencil-perf}), however, the speedup is even larger on the small tile sizes: on 32x32, it is 1.14x and 1.17x for the horizontal and vertical halos, respectively.
MPI messages between nodes are much more expensive compared to on-node, and the number of LDs per MPI operation is small, so the CXL LD overhead is not so substantial, while the constant \texttt{MPI\_LAT} dominates for small halo arrays.
For tile sizes 128x128 and above, we see almost no change in the expected performance.

These results predict that adopting message-free CXL data exchange on all halos would result in up to 1.37x speedup for the smallest tile size, while users see only marginal benefits on large halos, which discourages them from refactoring.
Using the upper-end \SI{300}{\nano\second} \texttt{CXL\_LAT} from~\cite{ahn2024mpi} and \SI{350}{\nano\second} \texttt{CXL\_ATOMIC\_LAT}, the speedup even grows to 1.59x.
Our results 1)~indicate that CXL carries significant potential to improve performance of different applications, especially in strong-scaling scenarios where cross-node communication costs dominate, and 2)~provide clear hints to the users on where to focus their valuable time to maximize performance.

\subsection{HPCG benchmark}\label{sec:hpcg}

The second  use case is the popular High Performance Conjugate Gradient (HPCG) benchmark, which complements High Performance Linpack (HPL) in the Top500 list.
HPCG reflects computation and data access patterns typical of real scientific applications, providing a more realistic workload for HPC systems.
It works on sparse data structures, and is more communication heavy and less compute intensive compared to HPL~\cite{heroux2013toward}.

Analog to the first use-case, we focus on the point-to-point communication (\texttt{MPI\_Send} and \texttt{MPI\_Irecv}), which is used to exchange boundary conditions between neighbors. 
Since HPCG manages data exchange with all neighbors in one loop, we analyze 3 options: 
\begin{itemize}
    \item Baseline MPI
    \item Optane (all neighbors)
    \item DDR (all neighbors)
\end{itemize}

Implementation details of HPCG introduce slight differences between MPI and shared memory versions.
The MPI implementation packs scattered data into a contiguous buffer before sending.
The shared memory buffer is written directly into the shared memory array.
Moreover, MPI does not require unpacking, as data is received directly into the Vector values, although only part of the buffer is used to receive data from other ranks.
CXL cannot allocate only a part of the buffer, while the rest stays in DDR, necessitating an additional unpack operation from CXL to the DDR buffer.
This change alters the data access pattern of the CXL version -- the unpack is a streaming access to CXL, and after that, the values are loaded again from DDR -- same as the baseline MPI version.
This illustrates that unpacking may be unavoidable in real-world applications.
Therefore, we only model the unpack from CXL (cf. Sec.~\ref{sec:data-access-overhead}).

We run HPCG in its default configuration, altering problem sizes, from 16-16-16 to 256-256-256.
Fig.~\ref{fig:hpcg-perf} presents the results of the reference implementation and the model predictions.
We observe several trends, many analogous to the previous use case:
(1) In the reference implementation (Fig.~\ref{fig:hpcg-real}), the relative differences in performance are smaller compared to the previous use-case, to a large extent because there is comparatively less communication between the ranks.
(2) The differences between the options become smaller with growing problem size, because communication becomes less dominant.
(3) The DDR implementation consistently outperforms Optane
(4) The MPI baseline is the best performing option in most cases.

Validation against model predictions (Fig.~\ref{fig:hpcg-model}) shows strong alignment:
(1,2) The impact of communication decreases with growing problem size, and (3,4) Optane remains slower, while DDR approaches MPI performance.
Our reference implementation dynamically allocates new shared memory buffers for each data transfer -- a simple and clean but inefficient solution that incurs a small overhead not accounted for in our model. \footnote{Unlike the 2D stencil code in Sec.~\ref{sec:stencil}, HPCG MPI buffers vary in sizes across levels, requiring our reference implementation to allocate a new buffer per transfer. Production-ready CXL data exchange implementation will certainly adopt a more efficient approach, but this is beyond the scope of this study.}
This provides an explanation why model predictions are 17 and \SI{8}{\milli\second} lower in the 16-16-16 size.
Despite this minor discrepancy, which may appear large on the plot due to the short runtime and small performance differences (max.~\SI{3}{\percent}), our results demonstrate the model's robustness in capturing overall trends.

\begin{figure}[t]
  \centering
  \includegraphics[width=0.49\textwidth]{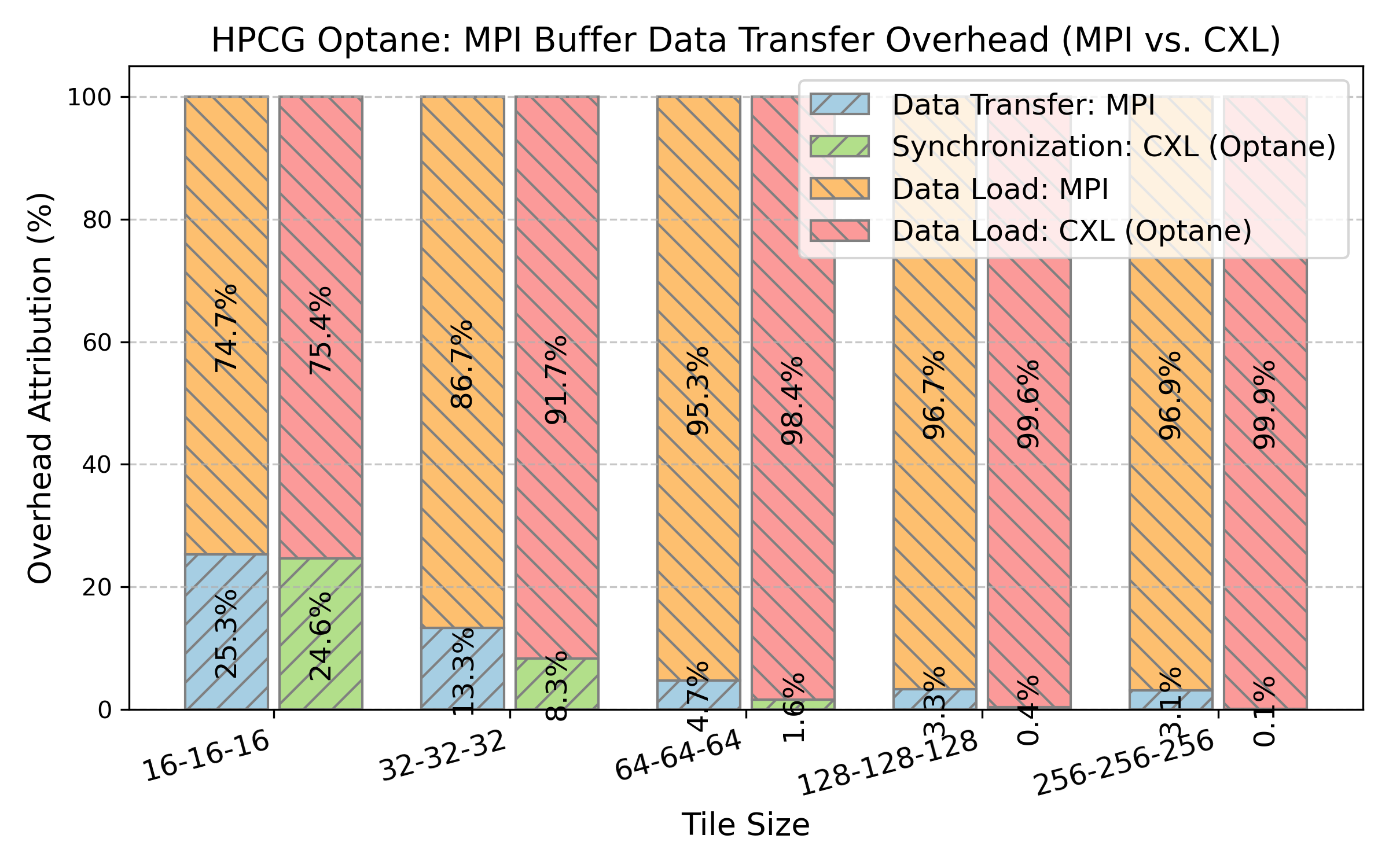}
  \caption{Breakdown of data load and transfer overhead for MPI and CXL (Optane) implementations in HPCG benchmark.}
  \label{fig:hpcg-breakdown}
\end{figure}

\subsubsection{Overhead Breakdown}
Fig.~\ref{fig:hpcg-breakdown} reveals a significant shift in overhead contributions, similar to observations in Sec.~\ref{sec:stencil-overhead-breakdown} and Fig.~\ref{fig:stencil-breakdown}.
In the Optane shared memory model, the data transfer overheads are up to \SI{25}{\percent} -- a considerable amount -- of the overall cost.
For MPI, transfer overhead drops to \SI{3}{\percent} for the largest size, leaving data loads as the dominant factor.
CXL data transfer overhead falls just to \SI{0.1}{\percent} -- significantly lower than MPI -- as the overhead does not depend on the buffer size.

Along with the results from Sec.~\ref{sec:stencil-overhead-breakdown}, we see data loads contributing anywhere between \SI{3.7}{\percent} (negligible) and \SI{99.9}{\percent} (dominant) of the overhead for different applications and problem sizes.
These observations highlight the \textbf{necessity to consider both the data transfer overhead and the data access overhead for a precise performance estimation} when comparing a baseline MPI with future CXL-based data exchange.
Despite its importance, the data access pattern analysis was not considered in prior MPI-over-CXL modeling studies, such as~\cite{ahn2024mpi,tran2024omb}.

%%%%%%%%%%%%%%%%%%%%%%%%%%%%%%%%%%%%%%%%%%%%%%%%%%%%%%%%%%%%%%%%%%%%%%%%%%%%%%
\section{Limitations and Discussion}

This work introduces a novel performance model that evaluates both data access and transfer overheads, along with an automated workflow for seamless application analysis and model prediction generation.
However, several limitations exist in model scope and validation.
First, due to the unavailability of CXL.mem hosts (expected in 2026), validation is restricted to a single-node setup -- the only feasible alternative for sharing a cache-coherent memory window between MPI processes.
Second, our analysis in this work targets point-to-point communication. 
Prior studies predict even greater speedups for collective opercommunicationations (e.g., over 40× for \texttt{MPI\_Allgather} in~\cite{ahn2024mpi}).
Our model can apply the network overhead model of these studies in the future; the data access overhead is already in place.
Finally, the Hockney-based network overhead model, which provided solid results in our experiments, does not account for network topology. 
Incorporating advanced models, such as those from the LogP* family, could enhance prediction accuracy for certain workloads.

%%%%%%%%%%%%%%%%%%%%%%%%%%%%%%%%%%%%%%%%%%%%%%%%%%%%%%%%%%%%%%%%%%%%%%%%%%%%%%
\section{Conclusions}

Upcoming CXL.mem-based systems will introduce memory pooling across multiple nodes, enabling shared access to cache-coherent memory segments.
This capability creates new opportunities for cross-node communication, positioning future systems between traditional shared-memory and shared-nothing architectures.
However, determining which MPI communications should transition to direct, message-free CXL.mem access remains non-trivial.

To address this challenge, we proposed a novel performance analysis and modeling workflow that predicts potential benefits (or drawbacks) of replacing MPI-based messaging with message-free CXL.mem communication.
Our approach combines both the actual communication and the load/store overheads from/to the communication buffers.
Our experiments revealed that either data transfer or data access overhead can dominate, depending on application characteristics and problem size, emphasizing the need for a detailed cost analysis of both aspects.
This was achieved through MPI tracing and hardware sampling using an extended Mitos toolset.
Per-MPI-call granularity enables developers to identify high-impact optimizations, improving performance while minimizing refactoring effort.

We showcased and validated the accuracy of the model predictions on a 2D stencil code, used to accentuate the importance of both communication and data access overhead.
As a second use-case, we presented results on the HPCG benchmark, comparing them with a reference on-node shared memory implementation.
Both use-cases provide correct guidance to the developer.
We further generated predictions for a multi-node run, which showed a significant performance benefit we can expect from a real CXL.mem setup, but also showed the limitations, highlighting the necessity to understand and carefully optimize applications to get the best overall system performance, combining traditional messaging and novel CXL.mem message-free communication.

%%%%%%%%%%%%%%%%%%%%%%%%%%%%%%%%%%%%%%%%%%%%%%%%%%%%%%%%%%%%%%%%%%%%%%%%%%%%%%

\bibliographystyle{IEEEtran}
\bibliography{sample-base}
\end{document}